\documentclass[twocolumn,showpacs,preprintnumbers,amsmath,amssymb]{revtex4}
\usepackage[dvips]{epsfig}
\usepackage{graphicx}
\usepackage{dcolumn}
\usepackage{bm}

\begin{document}


\title{Emergence of chaotic attractor and anti-synchronization for two coupled monostable neurons }

\author{M. Courbage$^{a}$}
\email{courbage@ccr.jussieu.fr}
\author{V.B. Kazantsev$^{b}$}
\author{ V.I.
Nekorkin$^{b}$}
\author{M. Senneret$^{a}$}
\affiliation{ (a) Universit\'e Paris 7-Denis Diderot/L.P.T.M.C.,
F\'ed\'eration Mati\`{e}re et syst\`{e}mes Complexes, 4
Place Jussieu, 75251 Paris Cedex 05, France\\ (b) Institute of Applied Physics of RAS,\\
Nizhny Novgorod State University, Nizhny Novgorod, Russia\\}

\date{\today}

\begin{abstract}

The dynamics of two coupled piece-wise linear one-dimensional
monostable maps is investigated. The single map is associated with
Poincar\'e section of the FitzHugh-Nagumo neuron model. It is
found that a diffusive coupling leads to the appearance of chaotic
attractor. The attractor exists in an invariant region of phase
space bounded by the manifolds of the saddle fixed point and the
saddle periodic point. The oscillations from the chaotic attractor
have a spike-burst shape with anti-phase synchronized spiking.

\end{abstract}

\maketitle

\newpage

\textbf{Spiking-bursting activity is a common feature of the
temporal organization in many neural firing patterns \cite{john}.
Bursting activity means that clusters of spikes (of action
potentials) occur more or less rhythmically and separated by
phases of quiescence. Spiking-bursting dynamics can be regular or
chaotic depending on the concentration of neuromodulators,
currents and other control parameters. For example, many of the
talamocortical neurons from central patterns generators generate
chaotic spiking-bursting dynamics. In order to understand the
emergence of chaotic oscillations in a neurones network, we use a
two variables FitzHugh-Nagumo model of the membrane potential  of
an isolated neural cell. For appropriate values of the parameters,
the system possesses a stable fixed point a focus surrounded by a
separatrix such that for large enough perturbations, the system
responds by an excitation pulse which also decays to the stable
focus. Hence, introducing a section by a half-line, we obtain a
one-dimensional piece-wise smooth  map with one discontinuity
point representing the excitation threshold. Then we consider  a
difference coupling between two such maps. In the linear
approximation of the maps and for suitable values of the
parameters, this two-dimensional system is locally hyperbolic
having discontinuity lines and a strange attractor . The
interesting fact is that these lines define regions in the phase
space corresponding to the thresholds of excitability of each one
of these neurones, or none of them. The symbolic dynamics
represents then the bursting activity of the neurones. It is then
possible to apply the approximation tools of statistical analysis
of time series generated by the system, namely through
Perron-Frobenius Operator, in order to study and interpret some
important features of neural networks, like synchronization and
time correlations.}

\section{Introduction}

Many neural systems composed of a number of interacting neurons
exhibit self-sustained oscillatory behavior leading to the
formation of various space-time patterns. Such patterns are
believed to play a key role in signal and information processing
functions of the brain \cite{kan,mur,sco}. One of the fundamental
problems is the understanding of possible dynamic mechanisms of
such patterns to appear and to evolve in time and space
\cite{man,loe,sher,wang1,rul,vries,kop,pinto}. There are two basic
phenomena of emergence of oscillations (regular or chaotic) in
neuron assemblies. The first one is obvious and deals with the
presence of local intra-cellular oscillations. Being coupled such
units are capable of various oscillatory patterns, clustering and
synchronization. The second one, found in recent theoretical and
experimental studies, concerns the possibility of oscillations in
assemblies of non-oscillatory cells \cite{man,loe,sher}. The
oscillations may also appear in coupled non-identical cells for
sufficiently strong coupling. The assembly is characterized by an
oscillatory ``average'' cell dynamics which makes the
non-oscillatory cells oscillating. Another studies have reported
that coupling even identical excitable cells can modify the
dynamics to form oscillatory attractors co-existing with a stable
fixed point \cite{sher}. Such attractors are characterized by
anti-phase spiking. The effects of electrical coupling between
neural cells also include the appearance of bursting in two
coupled pacemaker cells, modification of burst period in coupled
bursters, synchronization and chaos
\cite{wang1,rul,vries,kop,pinto,rul1}.

A model approach, in order to display the dynamical origin of
neural oscillations, is to use a simplified behavior-based
description of the system. For this purposes nonlinear maps could
be helpful as they can provide an appropriate qualitative
description of complex dynamic processes including chaotic
behavior in lower dimensional systems \cite{rul,vries,courb}. In
this paper we study the system of two coupled piece-wise linear
one-dimensional maps. The single map is derived from the dynamics
of an isolated neural cell modelled by the FitzHugh-Nagumo
excitable system \cite{sco,izh,kaz}. The FitzHugh-Nagumo neuron
model can be derived from Hodgkin-Huxley conductance-based model,
for some parameter values, when we take into account the
difference of kinetics between the \textit{potential dependent
gating variables} \cite{kan}. The first variable describes the
evolution of neuron membrane potential, the second mimics the
dynamics of outward ionic currents. Then, the model can describe
the salient features of neuron dynamics including the action
potential generation, excitability and excitation threshold.The
map has one globally stable fixed point and a discontinuity
corresponding to the excitation threshold of the cell. We shall
show how linear diffusive coupling between the two maps leads to
the appearance of chaotic oscillations with anti-phase spiking. A
number of studies of coupled chaotic maps have shown that
anti-phase chaotic oscillations may appear when the
synchronization manifold looses transverse stability and the
system evolves to the off-diagonal attractors
\cite{car,gle,ash,ash1,lai,courb}.

The paper is organized as follows. In Sect. II we show how the
dynamics of excitable FitzHugh-Nagumo model can be described by a
piece-wise continuous point map and introduce two-dimensional map
modeling a pair of coupled cells. In Sect. III we analyze the
dynamics of the map. We numerically show the emergence of strange
attractor in an invariant domain defined by invariant manifolds of
saddle fixed point and saddle periodic orbit. Sect. IV describes
the statistical characteristics of the attractor set and the
emergent chaotic oscillations with anti-phase spiking. Section V
contains a brief discussion of the results.

\section{Point map description of the excitable dynamics}

To replicate the excitable dynamics of an isolated neural cell one
can use the FitzHugh-Nagumo-like model. It can be taken in the
following form
\begin{equation}
\begin{array}{l}
\vspace{0.05 in} \dot{u}=u-u^3/3-v,\\
\vspace{0.05 in} \dot{v}=\epsilon (k(u)-v-I), \\
\end{array}
\label{fhn}
\end{equation}
The $u$-variable describes the evolution of the membrane potential
of the neuron, $v$ describes the dynamics of the outward ionic
currents (the recovery variable) \cite{sco}.The function $k$ is
taken piece-wise linear, $k(u)=k_1 u$, if $u<0$, and $k(u)=k_2 u$,
if $u \geq 0$. The parameters $k_1$ and $k_2$ control the shape
and the location of the $v$-nullcline, hence the dynamics of the
recovery term. The parameter $\epsilon$ defines the time scale of
excitation pulse and the parameter $I$ is a constant stimulus. The
excitable behavior of Eqs. (\ref{fhn}) is illustrated in Fig. 1.
Appropriate values of the parameters provide the existence of
three fixed points $O_1 (u^{(1)},v^{(1)})$, $O_2
(u^{(2)},v^{(2)})$ and $O_3 (u^{(3)},v^{(3)})$. The points $O_1$
and $O_3$ are stable and unstable foci, respectively, the point
$O_2$ is a saddle with the incoming separatrix defining the
excitation threshold. Then, if a perturbation of the rest state,
$O_1$, is large enough, i.e. lies beyond the separatrix, the
system responds with an excitation pulse, otherwise it decays to
the stable rest point $O_1$ (Fig. 1).

\begin{figure}
\includegraphics[width=\columnwidth]{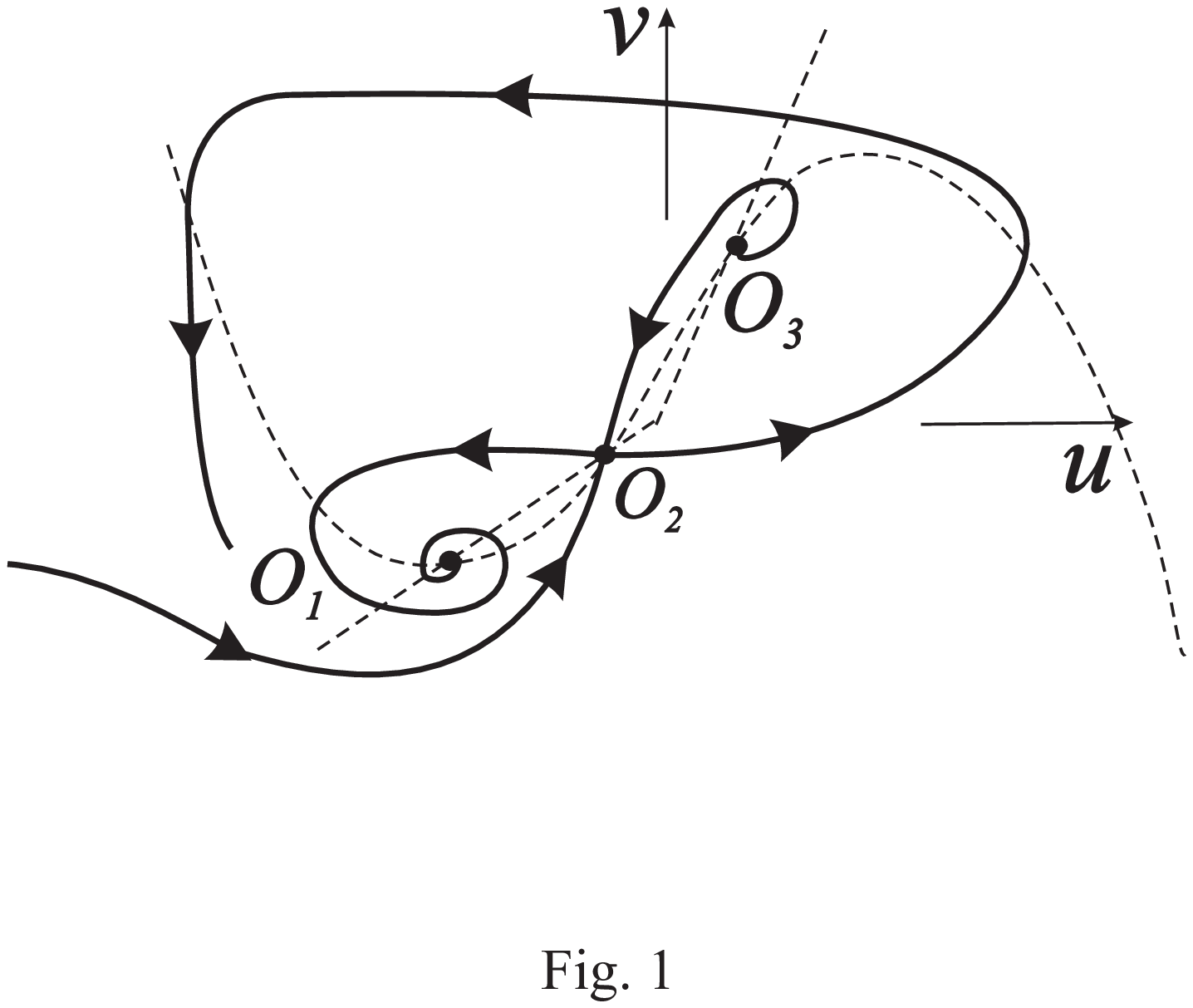}
\caption{Qualitative picture of the phase plane of Eqs.
(\ref{fhn}). $S$ illustrates the transversal half-line.}
\end{figure}

In order to describe the dynamics of the unit (\ref{fhn}) by a
point map, we introduce the transversal half-line $L: \{v=v^{(1)},
u>u^{(1)} \}$ (Fig. 1). It is found that the return map of the
flow given by Eqs. (\ref{fhn})  at the section $L$ defines a map,
$\Phi : L \rightarrow L$, for all points excluding the one, $L_0$,
corresponding to the intersection of the incoming saddle
separatrix with the half-line $L$. This point never returns to the
cross-section.  Consequently, we also must exclude all the
pre-images of the point $L_0$, $L_k=(\Phi^{-1})^m L_0$,
$m=1,2,\ldots, \infty$ . Then, the Poincar\'e map $\Phi$ can be
written as
\begin{equation}
x(n+1)=\Phi(x(n)), \label{map}
\end{equation}
$$ n=1,2,\ldots,\infty, $$
with $x(n)$ accounting for $(u-u^{(1)})$-coordinates of the points
at the Poincare section. The map is invertible and defined in the
interval $[0, \infty)$ excluding the points $L_k$ . The shape of
the curve $\Phi(x)$ calculated numerically is shown in Fig. 2. It
is given by a piece-wise continuous curve with one stable fixed
point. Then, the dynamics of the map is trivial. All its
trajectories represent sequences of points monotonically
approaching to the fixed point. The discontinuity point plays the
role of the excitation threshold, i.e. the neuron exhibits spiking
if the map evolves above this point. Note, that such map obtained
from (\ref{fhn}) gives an  ``average''  description of the cell
dynamics.

\begin{figure}
\centerline{\includegraphics[width=\columnwidth]{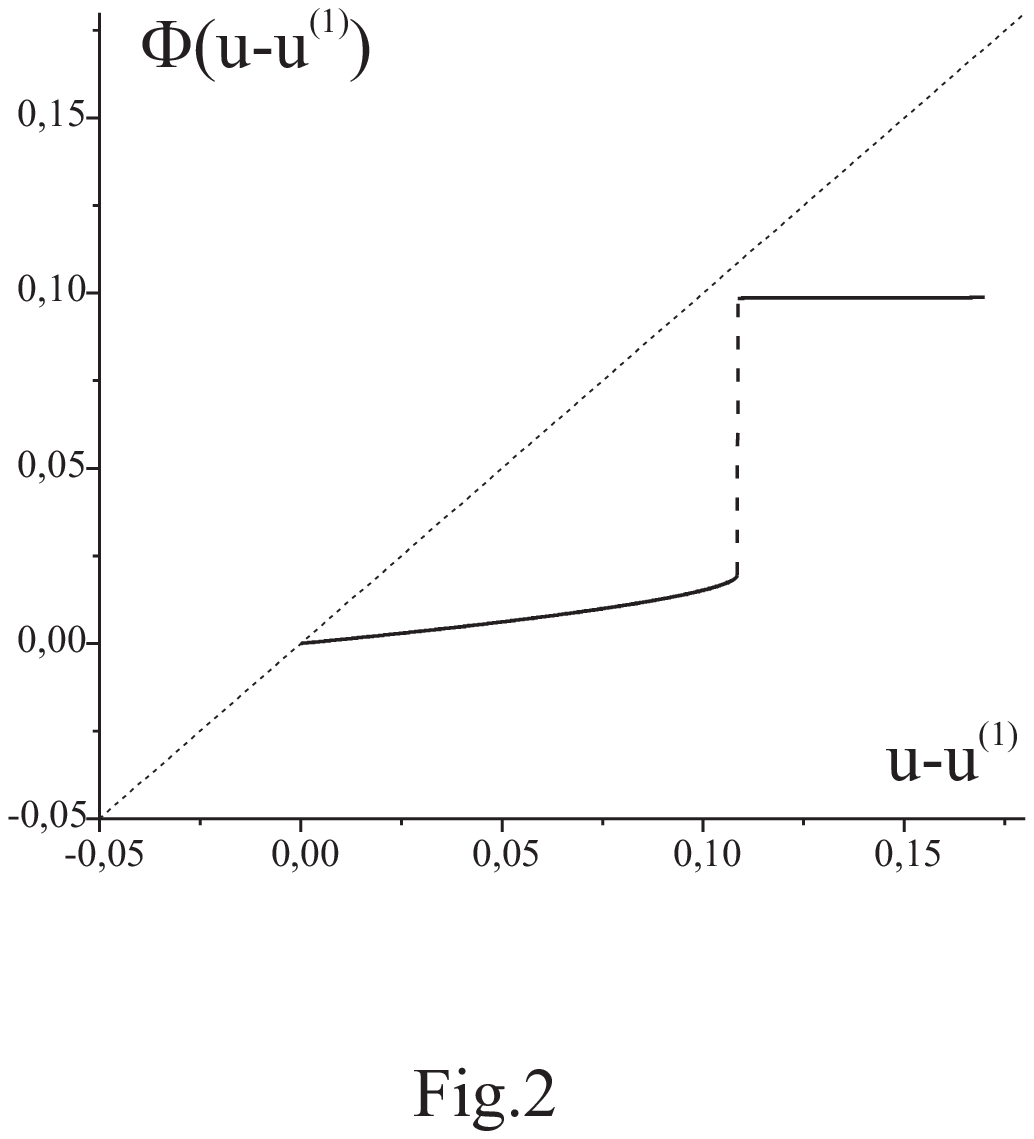}}
\caption{The curve $\Phi(x)$ obtained from Eqs. (\ref{fhn}) at the
section half-line $L$. Parameter values: $\epsilon=0.5, k_1=0.5,
k_2=2, I=0.221$.}
\end{figure}

Let us approximate the piece-wise continuous curve $\Phi(x)$ by a
piece-wise linear function and define the map for all points
$(-\infty, \infty)$ excluding the points $L_k$. Then, for two
cells we introduce difference coupling term (electrical coupling)
and consider the following two-dimensional map $f$: $R^2
\longrightarrow\ R^2 $
\begin{equation}
\begin{array}{l}
\vspace{0.05 in} x_1 (n+1)=F(x_1 (n)) + d (x_2 (n)-x_1(n)),\\
\vspace{0.05 in} x_2 (n+1)=F(x_2 (n)) + d (x_1 (n)-x_2(n)),
\end{array}
\label{2d}
\end{equation}
where the variables $x_{1,2}$ refer to the dynamics of the two
cells, $d$ is the coupling coefficient, the function $F(x)$ is
taken in the form
\begin{equation}
\label{fun} F(x)= \left\{
\begin{array}{l}
\alpha x, \; \mbox{if} \; x\leq a; \\
\alpha x +\alpha (b-a) \; \mbox{if} \; x>a,
\end{array}
\right.
\end{equation}

$0<\alpha<1$.

\section{Dynamics of the map}
\textbf{A. General properties of map}
\par\bigskip
\par\bigskip

Since $F(x)$ is given by the piece-wise linear function, then the
differential $Df$ of the map $f $ is a constant matrix, with
eigenvalues
\begin{equation}
\mu_1=\alpha - 2d, \; \mu_2=\alpha
\end{equation}.

Hence, Lyapunov exponents are defined by $\lambda_i=\ln|\mu_i |,
i=1,2$, for a arbitrary trajectory of the map, except zero
Lebesgue measure set of initial conditions.

Let us consider the system (\ref{2d}) in the parameter region
\begin{center}
$D=\{\alpha,d,a:\;0<\alpha<a/b,\;\frac{1+\alpha}{2}<d<1,\;a>0\}$
\end{center}

For convenience, we shall fixed $a=1$. This region $D$ is chosen
in order to have Lyapunov exponents $\lambda_1>0,\;\lambda_2<0$,
hence any trajectory, $\{f^i x_0,\; x_0\notin B\}$, where $B$ is a
set of discontinuous points of the map $f$, is heavily dependent
on initial conditions, and to have the modulus of Jacobian of the
map $f$ less than one, in $D$. All these features allow us to
speak of chaotic behavior.

We shall study the trajectories of the map $f$ missing the
discontinuity set $B=\bigcup\limits_{i=0}^\infty f^{-i} \gamma$,
where
\begin{equation}
\gamma=\{ x_1, x_2: \{ x_1=a \} \bigcup \{ x_2=a \} \}
\end{equation}

For convenience, let us change variables, according to the
eigenvectors of $Df$
\begin{equation}
y_1(n)=x_2(n)-x_1(n),\; y_2(n)=x_2(n)+x_1(n)
\end{equation}
Then, the discontinuity lines become
$$
\begin{array}{l}
\Gamma_1= \{y_1, y_2: \;y_2-y_1=2a\} \\
\Gamma_2=\{y_1,y_2:\;y_2+y_1=2a\}
\end{array}
$$

The lines $\Gamma_1$ and $\Gamma_2$ divide the phase plane
$(y_1,y_2)$ of the map $f$ into four regions

$$
\begin{array}{l}

G_1= \{y_1, y_2: \;y_2+y_1<2a, \;y_2-y_1<2a\} \\
G_2= \{y_1,y_2:  \;y_2+y_1>2a, \;y_2-y_1<2a\} \\
G_3= \{y_1, y_2: \;y_2+y_1<2a, \;y_2-y_1>2a\} \\
G_4= \{y_1, y_2: \;y_2+y_1>2a, \;y_2-y_1>2a\}

\end{array}
$$

In each region $G_i$ the map $f$ is continuous and has the form
\begin{equation}
\left\{
\begin{array}{l}
y_1(n+1)=\mu_1y_1(n)+b_1 \\
y_2(n+1)=\mu_2y_2(n)+b_2,
\end{array}
\right. \label{y1y2}
\end{equation}
with
$$
b_1= \left\{
\begin{array}{l}
0, \;if \;(y_1,y_2) \in G_1,G_4 \\
\alpha(b-a), \; if\; (y_1,y_2) \in G_2 \\
-\alpha(b-a),\;if\;(y_1,y_2) \in G_3
\end{array}
\right.
$$
$$
b_2= \left\{
\begin{array}{l}
0, \;if \;(y_1,y_2) \in G_1 \\
\alpha(b-a), \; if\; (y_1,y_2) \in G_2,G_3 \\
2\alpha(b-a),\;if\;(y_1,y_2) \in G_4
\end{array}
\right.
$$

The map $f$ has one hyperbolic fixed point, $O(0,0)$. In region
$G_1$ its invariant curves coincide with the coordinate axes.
Analyzing map $f$ in the regions $G_2$, $G_4$ we find that it has
a hyperbolic periodic orbit, $Q$, of period two with coordinates
\begin{equation}
\label{9} y_1=\frac{\alpha(b-a)}{2d-\alpha-1} \equiv y_1(Q),
\;\;\;\;\; y_2=\frac{\alpha(b-a)}{1-\alpha} \equiv y_2(Q),
\end{equation}
if the parameters satisfy the inequality
\begin{equation}
\label{10} d<d_p \;\;\; \mbox{with} \;\;\; d_p\equiv\frac
{2a-\alpha^2(a+b)}{2[2a-\alpha(a+b)]}.
\end{equation}

Stable and unstable invariant manifolds of the orbit $Q$ are
$$
W^s(Q)= \left\{y_1,y_2:\;
\begin{array}{l}
y_1=y_1(Q), \;\;\;\; if \; (y_1,y_2) \in G_2  \\
y_1=-y_1(Q), \; if \; (y_1,y_2) \in G_3 \\
\end{array}
\right.
$$
$$
\;\;W^u(Q)= \{\;y_1,y_2: \;\; y=y_2(Q), \; if \; (y_1,y_2) \in
G_2,G_2^-\}
$$

It follows from (\ref{9}) that
$$
\lim_{d\to\frac{1+\alpha}{2}+0} y_1(Q)=+\infty.
$$
Then periodic orbit $Q$ appears from infinity in the phase plane
$(y_1,y_2)$ for the parameters belonging to the curve
$$
N_{-1}=\left\{\alpha,d: \;\; d=\frac{1+\alpha}{2},
\;\;0<\alpha<\frac{a}{b} \right\}
$$
in the parameter plane $(\alpha,d)$. Note, that on this curve one
of the multiplier hits the bifurcation value,  $\mu_1=-1$. Then
fixed point $O$ changes its stability becoming of saddle type in
$D$.

\par\bigskip
\par\bigskip
\textbf{B. Invariant region and chaotic attractor of the map $f$}
\par\bigskip

Let us consider the location of invariant curves of the points $O$
and periodic orbit $Q$ in the phase plane with respect to the
lines $\Gamma_1,\;\Gamma_2$. The unstable invariant curve (the
separatix) $W^u_1(Q)$ intersects the line $\Gamma_2$ at point $K$
with coordinates
\begin{equation}
\label{11} y_1=\frac{2a-\alpha(b+a)}{1-\alpha}, \;\;\;\;\;
y_2=y_2(Q),
\end{equation}
and invariant curve $W^u_1(O)$ at the point $M$ with coordinates
\begin{equation}
\label{12} y_1=2a, \;\;\;\;\; y_2=0,
\end{equation}
Figure 3 illustrates the location of the invariant curves of fixed
point $O$ and orbit $Q$ in the phase plane $(y_1,y_2)$.

Let us introduce region $\Pi$ in the phase plane defined by
$$
\Pi=\left\{y_1,y_2: \;\; -y_1(Q)<y_1<y_1(Q), \;\;\; 0<y<y_2(Q)
\right\}
$$
Let us find the conditions for region $\Pi$ to be an invariant
region of map $f$. Since the system (\ref{y1y2}) is symmetric with
respect to $y_2$-axis it is sufficient to consider only $y_1\geq0$
region. Assuming ${(y_1(0),y_2(0))\in G_1^+}$, where ${G_1^+ =
\{\Pi\setminus B\} \bigcap G_1 \bigcap \{y_1\geq0\}}$, let us
obtain the conditions on the parameter values for which
${(y_1(1),y_2(1))\in \Pi}$. In region $G_1^+$ we find that
\begin{equation}
\label{13}
\left\{ \;
\begin{array}{l}
y_1(1)=-(2d-\alpha)y_1(0)  \\
y_2(1)=\alpha y_2(0)
\end{array}
\right.
\end{equation}
It follows from (\ref{13}) that ${(y_1(1),y_2(1))\in \Pi}$ if the
following inequalities
\begin{equation}
\label{14} -\frac{y_1(Q)}{2d-\alpha} < y_1(0)<
\frac{y_1(Q)}{2d-\alpha} \;\;\;\;\;\;\;\; y_2(0)<
\frac{y_2(Q)}{\alpha}
\end{equation}

are satisfied. Using the symmetry of $f$ with respect to $y_1$, we
can restrict the above condition to the right part of $\Pi$, and
since
\begin{equation}
\begin{split}
\max_{{(y_1(0),y_2(0))\in G_1^+}} y_1(0)& <  y_1(M),\\ \max
_{{(y_1(0),y_2(0))\in G_1^+}} y_2(0) & <  y_2(Q)
\end{split}
\end{equation}

the inequalities (\ref{14}) are satisfied if
\begin{equation}
\label{15} \frac{y_1(Q)}{2d-\alpha}>y_1(M), \;\;\;\;\;\;\;\;
\frac{y_2(Q)}{\alpha}>y_2(Q)
\end{equation}
The first inequalities in (\ref{15})impose the following condition
on $d$:
\begin{equation}
\label{16} d<d_h \;\; \mbox{with} \;\; d_h \equiv \frac
{1+2\alpha+\sqrt{1+\frac{2\alpha(b-a)}{a}}} {4}
\end{equation}

Thus, under the condition (\ref{16}) $f(G_1^+)\subseteq \Pi$.
Similarly we find that the image of ${G_2^+ = \{\Pi\setminus B\}
\bigcap G_2 \bigcap \;\{y_1>0\}}$ by $f$ is also included in
$\Pi$, for the parameter in $D$.

\begin{figure}
\centerline{\includegraphics[width=\columnwidth]{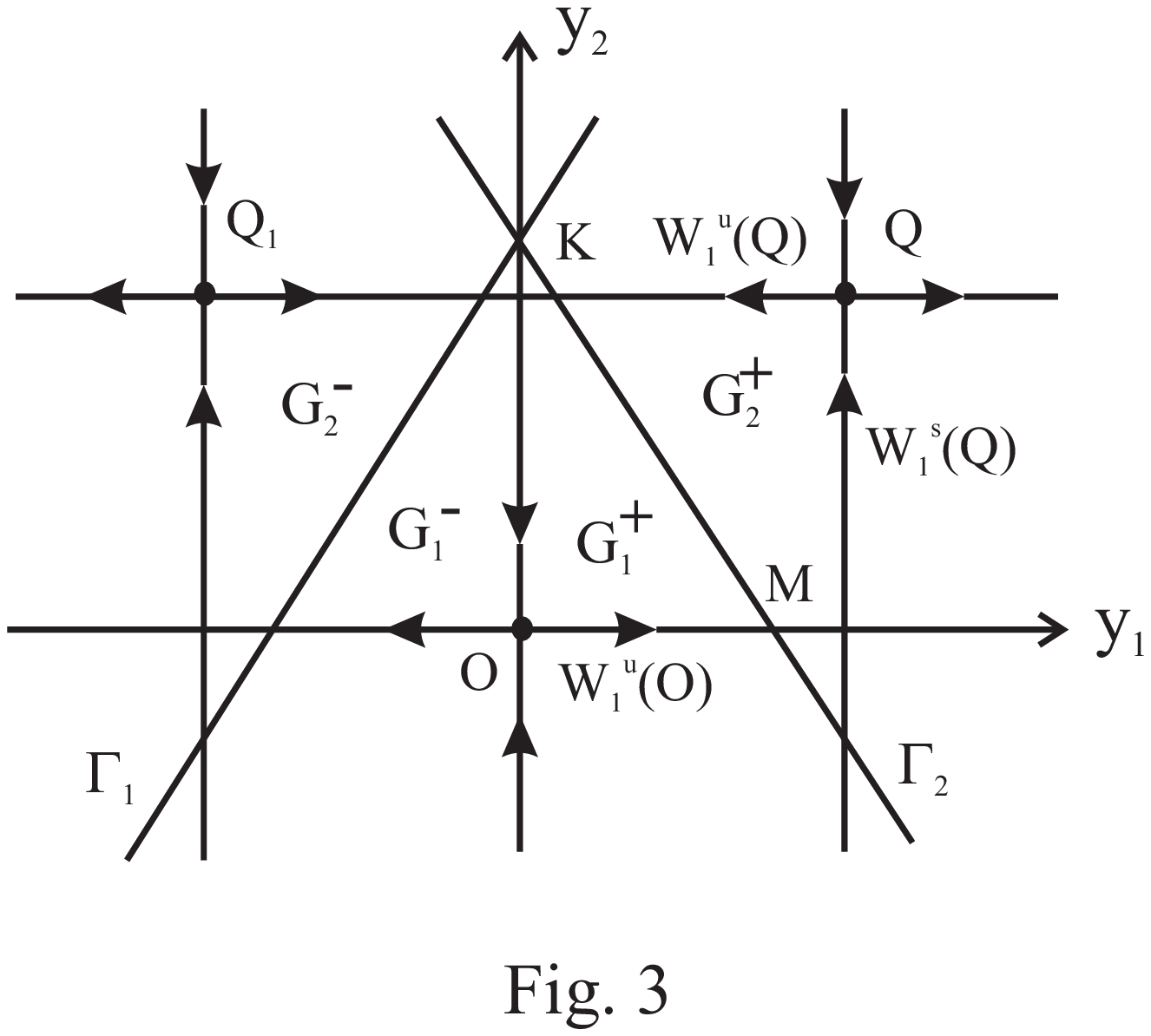}}
\caption{ Mutual location of invariant manifolds of the fixed
point $O$ and periodic orbit $Q$, and discontinuity lines
$\Gamma_{1,2}$ on the phase plane of map $f$.}
\end{figure}

Finally, $\Pi$ is invariant region of map $f$ in the parameter
region
$$
D_{inv}=D\bigcap \{d<d_p\} \bigcap \{ d<d_h\}
$$

\begin{figure}
\centerline{\includegraphics[width=\columnwidth]{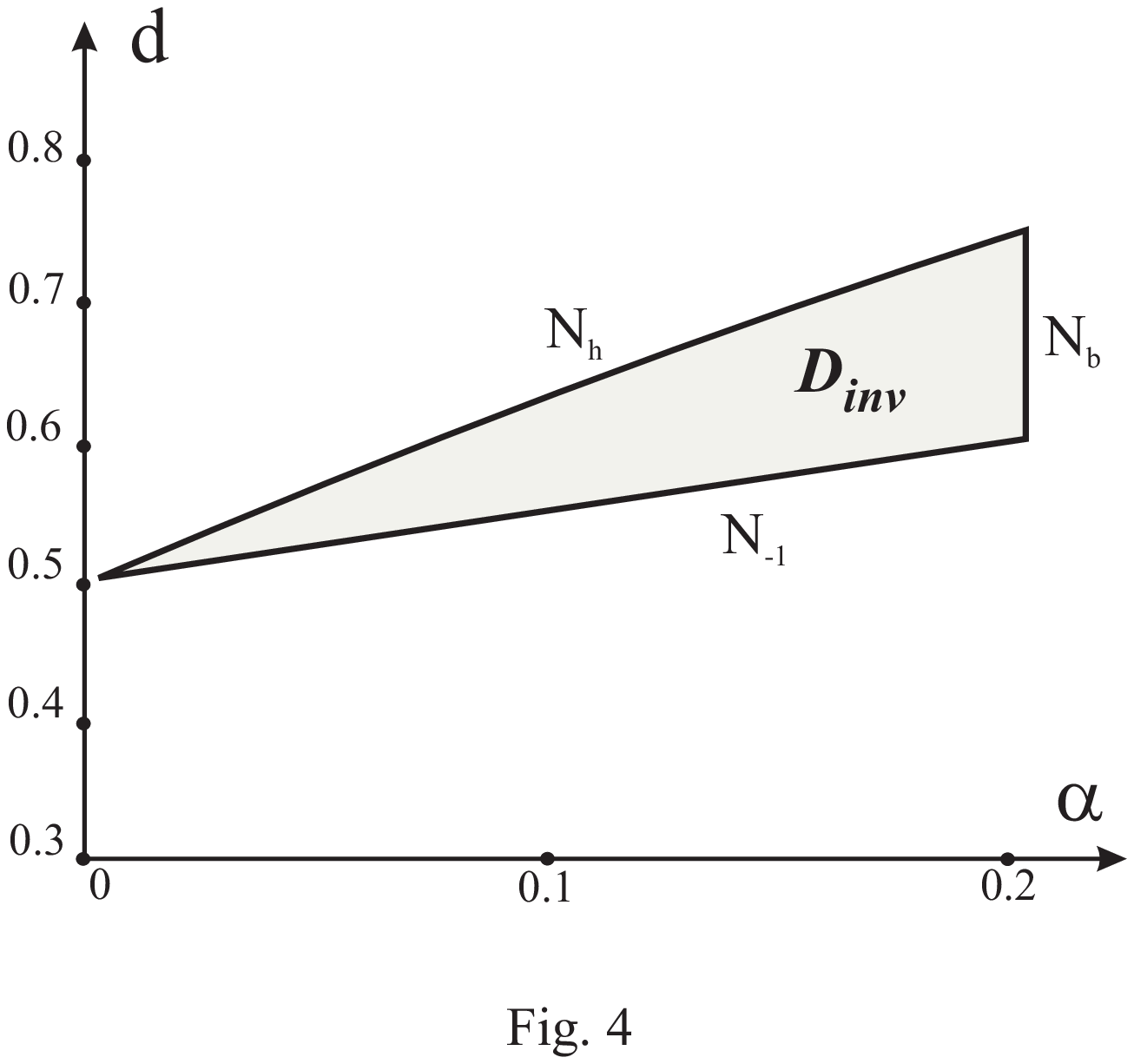}}
\caption{Parameter region $D_{inv}$ on the parameter plane
$(\alpha,d)$.}
\end{figure}

Figure $4$ shows region $D_{inv}$ in the parameter plane $(\alpha,
d)$. In this plane the boundary of $D_{inv}$ consists of three
components:
$$
\begin{array}{l}
N_{-1} = \left\{ \alpha,d: \; d=\frac{1+\alpha}{2}, \;
0<\alpha<\frac{a}{b} \right\} \\
N_b \;\; = \left\{ \alpha,d: \; \alpha=\frac{a}{b}, \;
\frac{a+b}{2b}<d<d_h \right\} \\
N_h \;\; = \left\{ \alpha,d: \; d=d_h, \; 0<\alpha<\frac{a}{b}
\right\}.
\end{array}
$$
The line $N_{-1}$ corresponds to the appearance of periodic orbit
$Q$ and changing stability of the fixed point $O$. The line $N_b$
is the boundary of the monostability of the uncoupled, $d=0$, maps
(\ref{2d}). The points of the curve $N_h$ corresponds to the
"tangency" of the separatricies $W_1^u(O)$ and $W_1^s(Q)$, and for
these parameter values $f(M) \in W_1^s(Q^-)$.

Therefore,  if the parameter values of  system (\ref{y1y2})
belongs to region $D_{inv}$, then invariant (absorbing) region
$\Pi$ and $\overline{f(\Pi \setminus B)} \subset \Pi$ exists in
the phase plane. Consequently, $\Pi$ contains strange (chaotic)
attractor $A$ of the map $f$. Figure 5 illustrates possible
structure of the attractor $A$ in the phase plane.

\begin{figure}
\centerline{\includegraphics[width=\columnwidth]{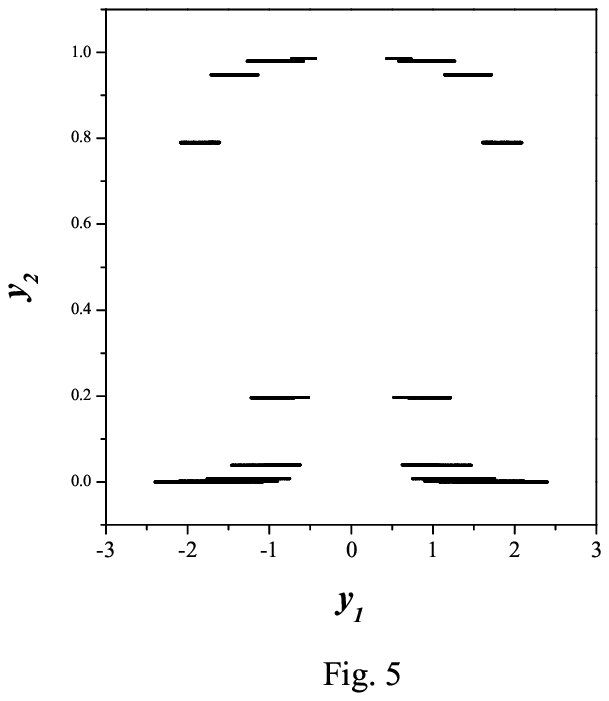}}
\caption{Chaotic attractor $A$ on the phase plane $(y_1, y_2)$.
Parameter values: $b=4.95, \alpha=0.2, d=0.74$.}
\end{figure}

To characterize the complexity of the chaotic attractor $A$ we
calculated numerically its fractal dimension $d_f (A)$.  It
appears that, indeed, $d_f(A)$ takes non-integer values greater
than $1$ (Fig. 6). Note that the box dimension of
the set increases with the increase of coupling coefficient $d$.
Corresponding estimate of the dimension $d_f$ using Lyapunov
exponents $\Lambda_i$ is shown by dashed curve.

\begin{figure}
\centerline{\includegraphics[width=\columnwidth]{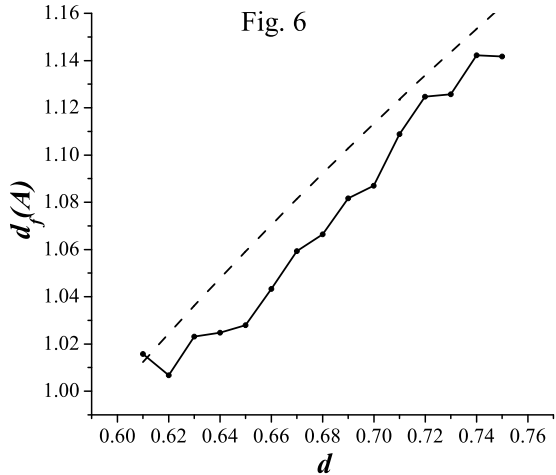}}
\caption{Fractal dimension $d_f(A)$ of the chaotic attractor $A$
calculated by box counting method versus coupling coefficient $d$.
Parameter values: $b=4.95, \alpha=0.2$.}
\end{figure}

\begin{figure}
\centerline{\includegraphics[width=\columnwidth]{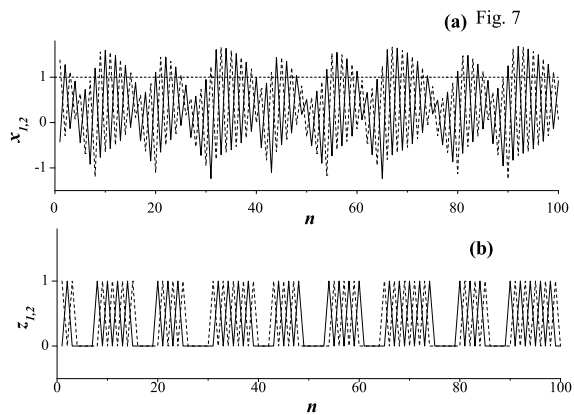}}
\caption{(a) Chaotic oscillations of the map (\ref{map}). $x_{1,2}
(n)$ are shown by
 solid and dashed curves, respectively. (b) Spike-burst behavior of binary variables $z_{1,2}
 (n)$. Parameter values: $b=4.95, \alpha=0.2, d=0.74$.}
\end{figure}

\par\bigskip
\par\bigskip
\textbf{C. Chaotic oscillations and attractor.}
\par\bigskip

Figure 7 (a) illustrates time evolution of the variables $x_1 (n),
x_2(n)$ corresponding to chaotic attractor $A$. Let us
characterize the oscillations in terms of original model
(\ref{fhn}) describing neuron excitability. Note that the neuron
excitation threshold accounted by saddle separatrix (Fig. 1)
corresponds to the discontinuity lines $\Gamma_1, \Gamma_2$ in the
map description. Thus, if a map trajectory jumps above these lines
then we can refer this event as a neuron spike, if it evolves
below $\Gamma_1, \Gamma_2$, the neuron is not excited . In such a
way the map oscillations can be described with binary variables
$z_i$:
$$
z_i (n) = \left\{
\begin{array}{ll}
0, \mbox{if} & x_i (n)<a; \\
1,  \mbox{if} & x_i (n)>a.
\end{array}
\right.
$$

The evolution of $z_i$-variables is shown in Fig. 7 (b). It
appears that map oscillation represent sequence of bursts
containing anti-phase spiking. Note that characteristic time
scales of oscillations (burst duration and inter-burst period)
depend coupling $d$. For smaller $d$ one can find longer lasting
subthreshold period and shorter burst period. Then, the map
describes a kind of chaotic spike-burst behavior typical for many
neural systems.

\section{Stationary probability distribution on the chaotic attractor}

The global behavior of the chaotic dynamics is described by the
probability of finding the trajectory in any given region of the
attractor. It can be visualized by a distribution of a cloud of
points, each moving under the deterministic mapping. A stationary
distribution may be reached by the system after long term
evolution. To obtain this distribution, we start from a histogram
of clouds of points in each small ``box'' of the phase space.
These points are mapped by $f$ and a new cloud is thus obtained.
Thus to the initial probability density $\rho_0$ correspond a new
density $\rho_1$. The operator which maps $\rho_0 \longrightarrow
\rho_1$ is called the Perron-Frobenius operator: $\rho_1=P\rho_0$
defined by:
\begin{equation}
\label{17}
\int_A P \rho_0 (y_1,y_2)dy_1dy_2=\int_{f^-1(A)} \rho_0
(y_1,y_2)dy_1dy_2
\end{equation}
for any region $A$ of the phase space. The stationary density
$\rho$ is an eigenfunction of $P$ corresponding to the eigenvalue
$1$. The existence and uniqueness of the stationary distribution
has been the object of many works \cite{boya,wang1}. It is
nevertheless difficult to obtain analytical exact expression of
$\rho(y_1,y_2)$, apart from the restricted case of Markov maps. We
shall use an approximating algorithm inspired from the method of
Ding and Zhou \cite{ding}.

The phase space of the coupled system is divided into $n \times m$
identical rectangles $\Delta_i$. We consider an initial
probability density $\rho_o (y_1,y_2)$ that is constant on each
$\Delta_i$:
\begin{equation}
\label{18} \rho_0(Y)= \sum_{i=1}^{n\times m}
\frac{a_i(0)}{m(\Delta_i)} \chi_{\Delta_i}(Y)
\end{equation}
where $Y=(y_1,y_2)$, $m(\Delta_i)$ is the Lebesgue measure of
$\Delta_i$ and $a_i(0)$ is the probability of finding a phase
point in $ \Delta_i , \; \chi_{\Delta_i}$ is the characteristic
function of $\Delta_i$:
$$
\chi_{\Delta_i} (Y)= \left\{
\begin{array}{l}
1 \;\;\;\; if \;\;\; Y \in \Delta_i \\
0 \;\;\;\; if \;\;\; Y \notin \Delta_i \\
\end{array}
\right.
$$
It is clear that $\rho_1$ has no reason to be of the
``Coarse-grained'' form (\ref{18}). Smoothing this density by
integrating it on each $\Delta_j$ we obtain by using the
definition (\ref{17}):
\begin{equation}
\label{19}
\begin{array}{rcl}
a_j(1)&=&\int_{\Delta_j}   P \rho_0 (Y) dY \\
&=&\int_{f^{-1}(\Delta_j)} \rho_0 (Y) dY \\
&=&\int_{f^{-1}(\Delta_j)} [\sum_i \frac{a_i(0)}{m(\Delta_i)} \chi_{\Delta_i}(Y)]dY \\
&=&\sum_i \frac{a_i(0)}{m(\Delta_i)} \int_{\Delta_j} \chi_{\Delta_i}(Y)]dY \\
&=&\sum_i a_i(0) \frac{m(f^{-1}(\Delta_j) \bigcap \Delta_i)}{m(\Delta_i)}  \\
\end{array}
\end{equation}
Thus, the transition probability from $\Delta_i$ to $\Delta_j$ is
given by the stochastic matrix:
\begin{equation}
\label{20} p_{i,j}=m(f^{-1}(\Delta_j)/\Delta_i)\equiv
\frac{m(f^{-1}(\Delta_j) \bigcap \Delta_i)}{m(\Delta_i}))
\end{equation}
The matrix $p=(p_{i,j})$ is an approximate value of the
Perron-Frobenius operator $P$. It can be proved that $p$ converges
to $P$ as $|\Delta_i|\longrightarrow 0$ \cite{ding}. Eq.
(\ref{19}) can be written under the matrix form:
\begin{equation}
a(t+1)=a(t)p,\ t=0,\ldots
\end{equation}
where $a(t)$ is the row vector $(a_1(t),\ldots,a_N(t))$ with $N=n
\times m$. Thus, an approximate stationary probability is the row
eigenvector $\upsilon=(\upsilon_1(n),\ldots,\upsilon_N(n))$ such
that:
\begin{equation}
\upsilon=\upsilon \cdot p,
\end{equation}

We first calculate the matrix elements of $p$ and then we compute
the eigenvector $v$. To calculate $p_{i,j}$ we use (\ref{20}). In
each rectangle $\Delta_i$ we put $k \times l$ points uniformly
distributes - then we compute the number of point $(Y \in \Delta_i
)$ such that $f(Y)\in\Delta_j$, and we divide this number by the
number of points in $\Delta_i$. This provides $p_{i,j}$.

We used this method to approximate the stationary distribution for
$\alpha = 0,2$ and $d=0,75$ (Fig. 8) showing a density distributed
over some region, which is to be compared with the Fig. 5 where we
have used only one trajectory. It is to be noted that the last
figure shows the attractor generated by one trajectory whereas
Fig.8 shows the approximate density of the attractor.

\begin{figure*}
\centerline{\includegraphics[keepaspectratio=true,width=\columnwidth]{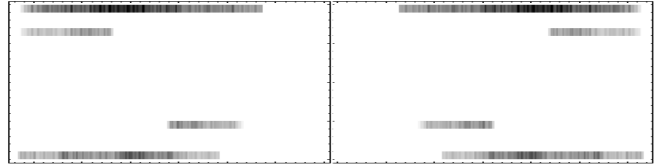}}
\caption{Approximative stationary distribution of the map
$f$.Parameter values: $b=4.95, \alpha=0.2, d=0.74$.}
\end{figure*}

\section{Synchronization.}

The above method allows to obtain the statistical distribution of
synchronized spikes. Recall that the variable $x_1$  is spiking if
and only if $Y\in G_2^-$ and $x_2$ is spiking if and only if $Y\in
G_2^+$, and no one is spiking if $Y\in G_1^+ \cup G_1^-$ which we
denote $G_1$. Recall also that the region $G_4$ corresponds to
$x_1$ and $x_2$ are simultaneously spiking, which lies outside the
invariant region. Moreover, because $G_1,\ G_2^{\pm}$ are
disjoint, $x_1$ and $x_2$ cannot be simultaneously spiking . Thus
we shall consider a partition of the invariant domain $\Pi$
corresponding to one of the three possible states of spiking
$S(Y)$:
$$
\begin{array}{rcl}
S(Y) = x_{1s} \;\;\; & \; if \; & Y \in G_2^- \\
S(Y) = x_{2s} \;\;\; & \; if \; & Y \in G_2^+ \\
S(Y) = 0 \;\;\;\;\;\; & \; if \; & Y \in G_1, \\
\end{array}
$$
where $x_{is}$ means that the only spiking variable is $x_i$; and
$0$ means that no one is spiking. This partition induces symbolic
dynamics, that we shall study. We calculate the stochastic matrix
$\pi_{i,j}$ from any one of these states to any other, where $i,j$
are the states $x_{1s},x_{2s},0$, for two values of the coupling
parameter $d=0.65,\ 0.7$ (see Fig. 9).
\begin{figure}
\centerline{\includegraphics[keepaspectratio=true,width=\columnwidth]{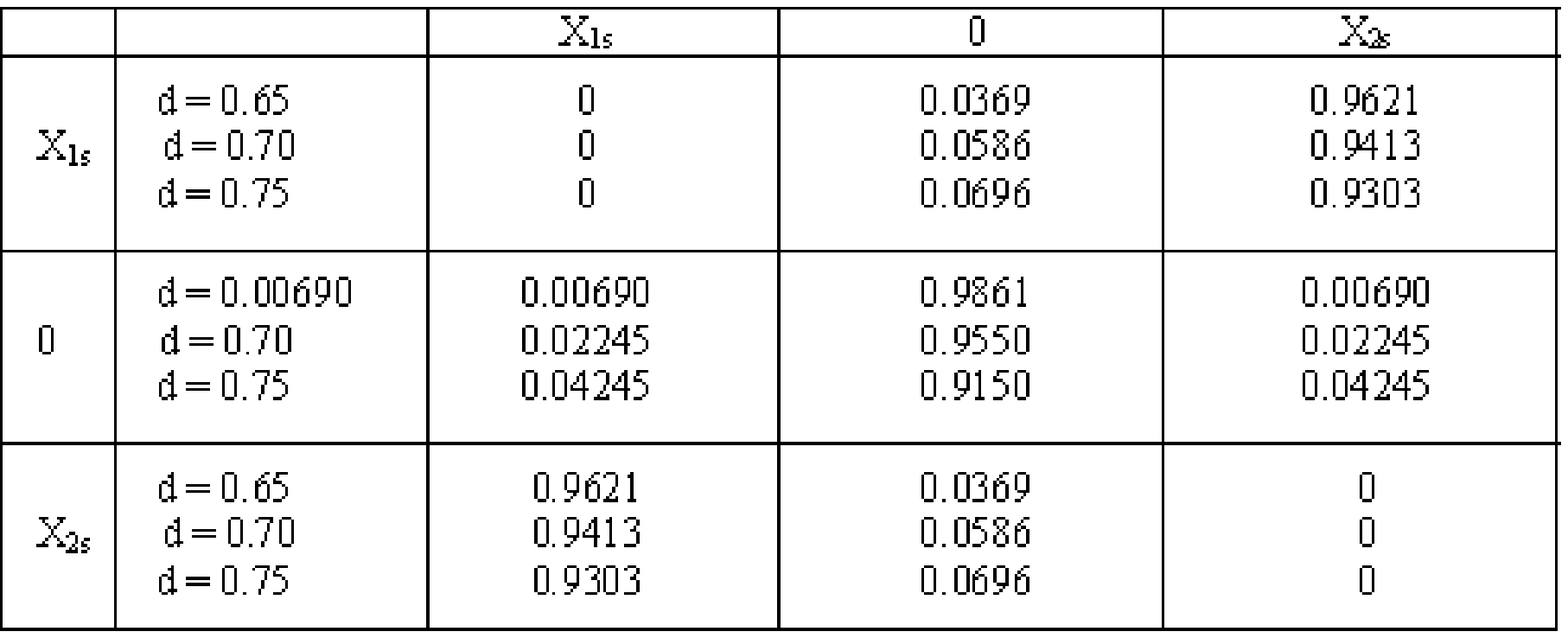}}
\caption{The $3*3$ stochastic matrices of transition probabilities
for three values of the coupling d.}
\end{figure}
This table shows that we have only the following probable
transitions in one step:
$$
x_{2s} \longrightarrow x_{1s} \longrightarrow x_{2s}
$$
and $0 \longleftrightarrow 0$.

Less probable transitions are:
$$
x_{1s}\longrightarrow 0, \;\;\; x_{2s}\longrightarrow 0
$$
$$
0
\begin{array}{ll}
& x_{1s} \\
\nearrow & \\
\searrow & \\
& x_{2s} \\
\end{array}
$$
but these transition probabilities depend monotonically on the
coupling $d$. Thus the transitions $x_{1s} \longrightarrow x_{2s}$
and $x_{2s} \longrightarrow x_{1s}$ (i.e. successive spikes)
decreases slightly with coupling. So, strengthening the coupling
in this range decreases the bursting probability.

We can understand these occurrences in considering the
intersections of the image of each region by $f$ with these same
regions (see Fig. 10). The most probable transition $x_{ls}
\longrightarrow x_{ks},\ \{k,l\} \in \{1,2\}^2 $ and $0
\longrightarrow 0$ correspond respectively to $f(G_2^{\pm})\bigcap
G_2^{\mp}$ and $f(G_1)\bigcap G_1$. The less probable transitions
$x_{ls} \longrightarrow 0$ and $0 \longrightarrow x_{ls}$
correspond respectively to $f(G_2\pm)\bigcap G_1$ and
$f(G_1)\bigcap G_2^{\pm}$. The positive Lebesgue measure of these
intersected regions give the transition probabilities of each of
these occurrences. Some of the transition probabilities given by
the calculation of the Lebesgue measures of the intersected
regions are shown in Fig. 11, with their dependance over the
coupling constant $d$, which can be compared to those obtained
numerically in the table Fig. 9. All the calculations are made in
appendix A, where we also show that ,when the parameters are in
$D_{inv}$,a neuron can not fire twice, i.e. $G_2^\pm\bigcap
f(G_2^\pm)=\emptyset$.

\begin{figure}
\centerline{\includegraphics[keepaspectratio=true,width=\columnwidth]{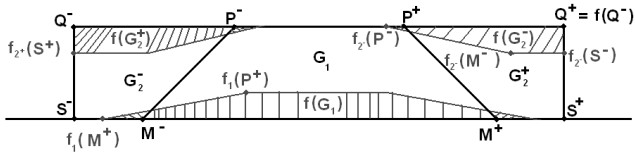}}
\caption{Image by $f$ of the partitions of the invariant region
$\Pi$.}
\end{figure}

\begin{figure}
\centerline{\includegraphics[keepaspectratio=true,width=\columnwidth]{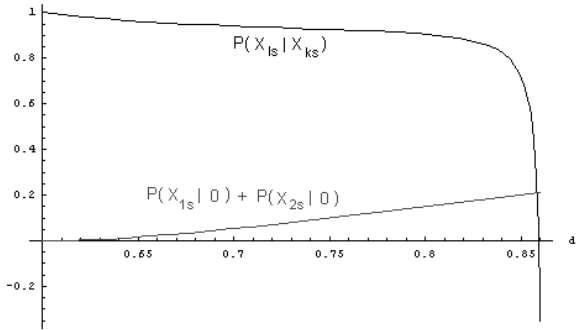}}
\caption{Probability transition from a spike to another, and from
rest to spike, versus coupling constant $d$; $b=4.95, \alpha=0.2$.
}
\end{figure}

We can now check the memory effects in studying the conditional
probability of a state given the past. If we denote by
$S_0,S_{-1},S_{-2} \ldots$ the values of the variable $S$ at time,
$0, \; -1, \; -2$ etc, we like to know if conditional probability
$Prob(S_0 | S_{-2},S_{-1})$ is equal to $Prob(S_0 \; | \; S_{-1})$
or not. In the first case there is no memory of the process. In
order to compute the above conditional probability we have to find
$Prob(S_0=i_0 \; | \; S_{-2}=i_{-2}, S_{-1}=i_{-1})$ which implies
$(3)^3=27$ calculations. But, neglecting rare events like
$S_{-2}=0, S_{-1}=x_{2s}$ having $0,0422$ probability for
$d=0,75$, we have only $4$ significant transitions.
$$
\begin{array}{l}
Prob(S_0=x_{2s} \; | \; S_{-2}=x_{2s}, \; S_{-1}=x_{1s})=P_1 \\
Prob(S_0=x_{1s} \; | \; S_{-2}=x_{1s}, \; S_{-1}=x_{2s})=P_2 \\
Prob(S_0=0      \; | \; S_{-2}=x_{2s}, \; S_{-1}=x_{1s})=P_3 \\
Prob(S_0=0      \; | \; S_{-2}=x_{1s}, \; S_{-1}=x_{2s})=P_4
\end{array}
$$

\begin{figure}
\centerline{\includegraphics[keepaspectratio=true,width=\columnwidth]{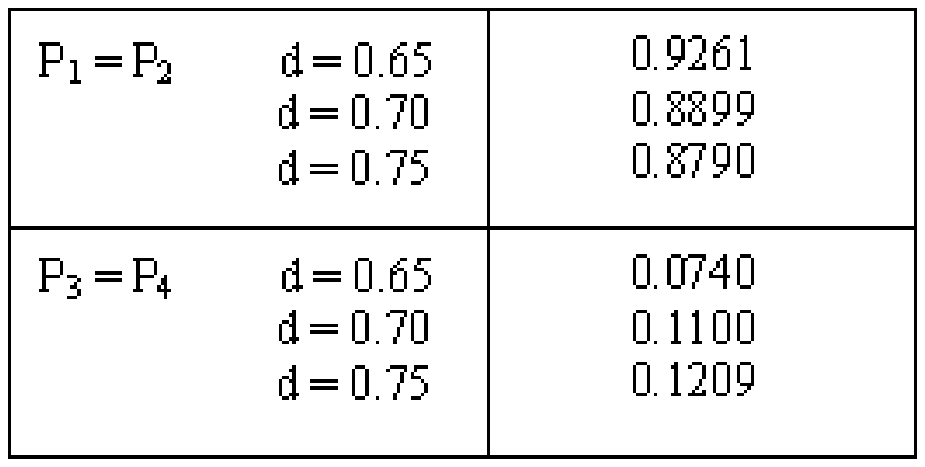}}
\caption{ The conditional probabilities of spiking and non-spiking
after two successive spikes respectively.}
\end{figure}

The first one is the probability of having $x_{2s}$  if it is
preceded by a sequence of $x_{2s}$ and $x_{1s}$ and for the others
the definition is the same. Thus we obtain the following table
(Fig.12). The fact that the probability $P_1=0,879$ is not equal
to $Prob(S_0=x_{2s} \; | \; S_{-1}=x_{1s}) = 0,9303$ means that
the system has acquired a memory, the chain is not a Markov chain.
The memory of a chain is given by the smallest positive integer
$m_o$ such that:
$$
Prob(S_0 \; | \; S_{-m}, \; S_{-m+1}, \ldots ,S_{-1})=Prob(S_0 \;
| \; S_{-m_0, \ldots, S_{-1}})
$$
for any $m \geq m_o$, and any value of $(S_0, \ldots S_{-m})$.

 It is also remarkable that the probability of having $0$
if it is preceded by the sequence $(x_{2s},  \; x_{1s})$ is
$0,1209$ the double of the probability of having $0$ preceded by
$x_{1s}$ simply $0,0696$. This explain the frequent appearance of
sequence of spikes. How could we estimate the memory of the
process depends on the full calculation of all conditional
probabilities for all past.

Another very interesting aspect is the dependence of this memory
on the coupling coefficient as shown in Fig. 13. It is clear that
the memory effects increase with coupling (the distance between
the two curves increases with coupling).

\begin{figure}
\centerline{\includegraphics[keepaspectratio=true,width=\columnwidth]{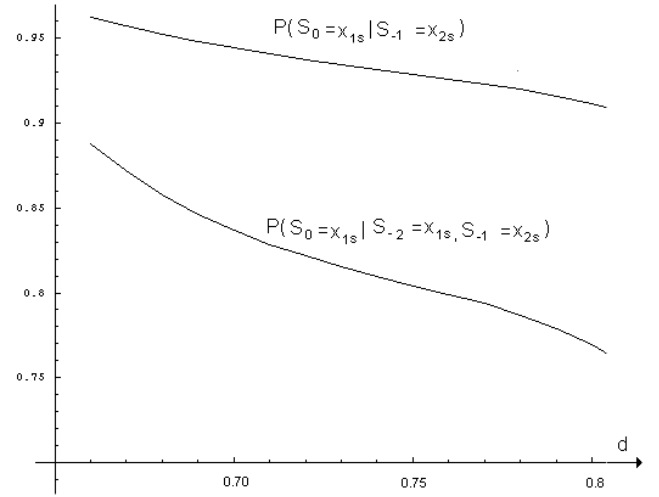}}
\caption{The effect of coupling on the memory effect. Parameter
values: $b=4.95, \alpha=0.2$.}
\end{figure}

\section{Conclusion}

We have investigated chaotic dynamics of two coupled maps each
formally derived from FitzHugh-Nagumo model of neuron
excitability. Being uncoupled such maps are trivial with only one
stable fixed point corresponding to neuron rest state. The
excitation threshold is given by the discontinuity point. We have
introduced the formal definition of spike if the map evolves above
this threshold. The linear ``diffusive'' coupling term introduced
for the maps can be treated as a kind of electrical inter-neuron
coupling that mimics an ``integral'' coupling current between the
two cells.

We have shown that increasing the coupling coefficient above
certain threshold leads to the appearance of chaotic attractor.
The attractor appears in the invariant region of the phase space
confined by the invariant manifolds of saddle periodic orbit. This
region attracts trajectories from outside and does not contain any
stable trajectory inside. The oscillations emerging with such an
attractor have a spike-burst shape with synchronous bursts with
anti-phase spiking. Using probabilistic description we have found
the probabilities to find a spike in different conditions. It has
been also found that the system acquires a memory (in contrast
with Markov's processes).

\section*{Acknoledgments}
This research has been supported by Russian Foundation for Basic
Research (grants 03-02-17135) and by grant of President of Russian
Federation (MK 4586.2004.2). V.B.K. acknowledges Russian Science
Support Foundation for financial support. V.I.N. acknowledges the
financial support from University Paris VII.

\appendix

\section{Image of the invariant region after one iteration.}

In order to understand the oscillatory behavior it is interesting
to look at the image of the invariant region after one iteration
(see Fig. 10). First of all, we have to determine the images of
the regions of the invariant region, defined by the
discontinuities, namely $G_1$, $G_2^+$ and $G_2^-$(e.q. to
$G_2^{-}$).

$G_1$ is a polygon having for vertices $P^+$, $P^-$, $M^+$ and
$M^-$; $G_2^+$, $Q^-$, $P^-$, $S^-$ and $M^-$; and $G_2^-$ ,
$P^+$, $Q^+$, $S^+$ and $M^-$ (when $M^+ < S^+$ or equivalent,
which is always satisfied for $d<d_h$). Let us compute the images
of all these vertex, according to the definition of $f$ in each
region. We have to keep in mind that some of these points belong
to the lines of discontinuity; in this case, the computation is
taken in the sense of a limit starting from the interior of each
region.

First, we compute the image of $G_1$ obtained by the image of its
vertices:

\begin{equation}
\begin{array}{l}
 f_1(M^{+})=\{ 2 a (\alpha - 2 d), 0\},\\
 f_1(M^{-})=\{-2 a(\alpha - 2 d), 0\},\\
 f_1(P^{+})=\{ { (\alpha (b+a)-2a)(\alpha-2 d)\over \alpha -1},{\alpha^2(a-b)\over \alpha-1}\},\\
f_1(P^{-})=\{ -{ (\alpha (b+a)-2a)(\alpha-2 d)\over \alpha
-1},{\alpha^2(a-b)\over \alpha-1}\},
\end{array}
\label{fG1a}
\end{equation}
we shall show that $(f(G_1)\bigcap G_2^+)\bigcup (f(G_1)\bigcap
G_2^-)\neq \emptyset$, which means that a neuron at rest can fire.
That is to say
\begin{equation}
\begin{array}{l}
y_1(f_1(M^-))>y_1(M^+),\\
-2a(\alpha-2d)>2a
\end{array}
\label{fG1b}
\end{equation}

However, the last inequality is always true when we are in
$D_{inv}$.

Next, we compute the image of $G_2^+$:

\begin{equation}
\begin{array}{l}
 f_2(M^{+})=\{ \alpha (-a + b) + 2 a (\alpha - 2 d), \alpha (-a + b)\},\\
 f_2(P^{+})=\{{ \alpha b (-2 \alpha +1+2 d)+a(\alpha+2d(\alpha-2))\over 1-\alpha}, {\alpha(b-a)\over 1-\alpha}\},\\
 f_2(Q^+)=\{ {\alpha (b-a)\over \alpha+1-2d},{\alpha (b-a)\over 1-\alpha} \},\\
f_2(S^+)=\{ { \alpha (b-a) \over\alpha+1-2d},{\alpha (b-a)}\},
\end{array}
\label{fG2a}
\end{equation}

We first shall see that $f(G_2^+)$ is completely included in
$G_1\bigcup G_2^-$ and excluded of $G_2^+$, which implies that one
neuron can not spike twice. For the simplicity of proof, we shall
use indirect calculations. We are going to show that:

\begin{equation}
\begin{array}{l}
y_1(f_2(M^+))<0 ,\\
y_1(f_2(K))>0,\\
y_2(f_2(K))>y_2(f_2(K)))\\
y_2(f_2(K))-y_1(f_2(K)< 2a
\end{array}
\label{fG2b}
\end{equation}
The first three inequalities allow us to say that $f_2(M^+)$ is
under and on the left of $f_2(K)$; but, as $P^+$ is on the segment
$[K\ M^+]$, $f_2(P^+)$ is on the segment $[f_2(K)\ f_2(M^+)]$, and
so it is under and on the left of $f_2(K)$.The last inequality
show that $f_2(K)$ (and by consequence$f_2(P^+)$)is not in $
G_2^+$.

First, we verify the first inequality
\begin{equation}
\begin{array}{l}
y_1(f_2(M^+))<0 ,\\
\alpha(b-a)+2 a(\alpha-2d)<0,\\
\alpha(b+a)<4 a d,
\end{array}
\label{fG2c}
\end{equation}

However in $D_{inv}$,
\begin{equation}
\begin{array}{l}
\alpha(b+a)<a+\alpha a,\\
2a +2 \alpha a < 4 a d,
\end{array}
\label{fG2d}
\end{equation}

but $a+\alpha a<2a +2 \alpha a$ is always satisfied, as $a>0,\\
\alpha>0$. By consequence,
$$y_1(f_2(M^+))<0.$$

The second and the third inequalities are straightforward as
$(b-a)>0$, $a>0$  and $\alpha>0$.

Finally, we check the last inequality:
\begin{equation}
\begin{array}{l}
y_2(f_2(K))-y_1(f_2(K)=\alpha(b-a)+2 a \alpha -\alpha(b-a),\\
y_2(f_2(K))-y_1(f_2(K)=-2\alpha b<0<2 a,
\end{array}
\label{fG2e}
\end{equation}
and so a neuron can not fire twice.

Finally, we compute $G_2^-$:
\begin{equation}
\begin{array}{l}
 f_3(M^{-})=\{ -\alpha (-a + b) - 2 a (\alpha - 2 d), \alpha (-a + b)\},\\
 f_3(P^{-})=\{{ \alpha b (2 \alpha -1-2 d)-a(\alpha+2d(\alpha-2))\over 1-\alpha}, {\alpha(b-a)\over 1-\alpha}\},\\
 f_3(Q^-)=\{ -{\alpha (b-a)\over \alpha+1-2d},{\alpha (b-a)\over 1-\alpha} \},\\
f_3(S^-)=\{ -{ \alpha (b-a) \over\alpha+1-2d},{\alpha (b-a)}\},
\end{array}
\label{fG2f}
\end{equation}
 By symmetry, the above results for $G_2^+$ hold for $G_2^-$.

 The computation of the Lebesgue measure of the regions of
 intersection involve only computation of polygonal area which it
 is straightforward with the coordinates of the vertices.

\end{document}